\documentclass[review]{elsarticle}
\usepackage[LGRgreek]{mathastext}

\usepackage{amssymb}

\journal{Nuclear Inst. and Methods in Physics Research, A}

\usepackage{HGTD_TDR-defs}
\usepackage{multirow} 
\usepackage{rotating}
\usepackage{mathrsfs}
\usepackage{booktabs}
\usepackage{longtable}
\usepackage{placeins}
\usepackage{amssymb}
\usepackage{upgreek}
\usepackage{svg}
\usepackage{comment}
\usepackage{pdfpages}
\usepackage{siunitx}
\usepackage{lineno}
\usepackage{float}
\usepackage{lineno}
\usepackage{tikz}
\usepackage[percent]{overpic}
\usepackage{caption}
\usepackage{subcaption}
\usepackage{makecell}
\usepackage{gensymb}
\usepackage[colorinlistoftodos]{todonotes}
\usepackage{lineno}
\title{\boldmath Extreme neutron fluence effects on AC-LGAD sensors}
 
\author[1]{A.~Esparza}
\author[1]{A.~Borjigin}
\author[1]{M.~Davis}
\author[1]{F.~McKinney-Martinez}
\author[1]{S.M.~Mazza$^*$}
\author[1]{A.~Molnar}
\author[1]{A.~Gsponer}
\author[1]{H.F.-W.~Sadrozinski}
\author[1]{B.~Schumm}
\author[1]{A.~Seiden}
\author[1]{T.~Shin}
\author[2]{G.~Kramberger}
\author[2]{I.~Mandic}

\affiliation[1]{SCIPP, University of California, Santa Cruz, 1156 High Street, Santa Cruz (CA), US}
\affiliation[2]{IJS, Jamova cesta 39, 1000 Ljubljana, Slovenia}

\usepackage{sfmath}

\begin{document}

\begin{frontmatter}

\begin{abstract}
LGAD and AC-LGAD devices from HPK were irradiated to extreme fluences at the TRIGA reactor in Ljubljana; the maximum fluence applied is 1e18~$n_{eq}/cm^2$.
The sensors were characterized before and after irradiation with a probe station (IV/CV) and a laser TCT setup. 
A series of test structures to measure the resistive layer resistivity and other wafer properties were irradiated as well and measured afterwards.
The devices were produced for the ePIC detector time of flight (TOF) layer, but the study is geared towards future applications since the radiation requirements for ePIC are several orders of magnitude less, only up to 1e13~$n_{eq}/cm^2$ fluence over the lifetime of the experiment. 
\end{abstract}

\begin{keyword}
fast silicon sensors; irradiation; charge multiplication; AC-LGAD strips; charge sharing.
$^*$Corresponding author: simazza@ucsc.edu
\end{keyword}

\end{frontmatter}


\section{Introduction}
\label{sec:intro}
Low Gain Avalanche Detectors (LGADs) have been established in the last 10 years as a fast semiconductor timing technology~\cite{Pellegrini:2014lki, Sadrozinski:2013nja} with a timing resolution of the order of tens of picoseconds. 
In LGADs' first large-scale experimental applications, the High Granularity Timing Detector (HGTD) in ATLAS~\cite{Mazza:2019dkn} and the MIP Timing Detector (MTD) in CMS~\cite{Ferrero:2022ynt}, the segmentation is limited to pads with about 1 mm pitch by consideration of power, fill-factor, and field uniformity. 
The fill-factor and uniformity are both solved for AC-LGAD technology, aka Resistive Silicon Detector (RSD)~\cite{BISHOP2024169478,Mandurrino}, which is based on a complete integration of four of the sensor layers in common sheets of the P-type bulk, the P$\textsuperscript{++}$ gain layer, the N$\textsuperscript{+}$ layer, and a dielectric sheet, separating the first three from the segmented metal readout contacts (Fig.~\ref{fig:ACLGAD}).
A signal originating in the bulk and amplified in the gain layer is then shared between several electronic channels, allowing signal location reconstruction with a resolution of a small ($<5\%$~\cite{MENZIO2024169526}) fraction of the readout pitch, also allowing an acceptable power density for high position and time resolution. 

AC-LGADs are the chosen sensor technology for the ePIC detector~\cite{AbdulKhalek:2021gbh} time-of-flight (TOF) layer, which is composed of strips in the barrel region and pixels in the end-cap region~\cite{davis2025characterizationfullsizeproductionepic}.
Strip and pixel AC$-$LGADs produced by Hamamatsu Photonics K.K. (HPK)~\cite{Hamamatsu} for ePIC were tested before and after extreme neutron irradiation with 1 MeV reactor neutrons at JSI/Ljubljana to fluences up to 1e18~$n_{eq}/cm^2$.
The radiation hardness requirement of ePIC TOF is only up to 1e13~$n_{eq}/cm^2$, therefore, this study utilizes detectors produced for ePIC but has the scope of general R\&D geared towards future applications like FCC-hh~\cite{FCCHH}.
The sensors were characterized with a probe station for current versus voltage (IV) and capacitance over voltage (CV) to probe the breakdown and degradation of the gain layer with irradiation.
Then, a selection of sensors was tested with a focused laser TCT station to test the change of behavior after irradiation in terms of charge-sharing response, time of arrival, and rise time. 
Previous studies showed almost no change in the behavior of AC-LGADs up to a fluence of 5e14~$n_{eq}/cm^2$~\cite{Stage:2025jre} except for the gain loss due to acceptor removal, which is similar to standard DC-LGADs.

\section{Sensors irradiated and tested}
\begin{figure}[!hbt]
 \centering
 \includegraphics[width=0.28\columnwidth]{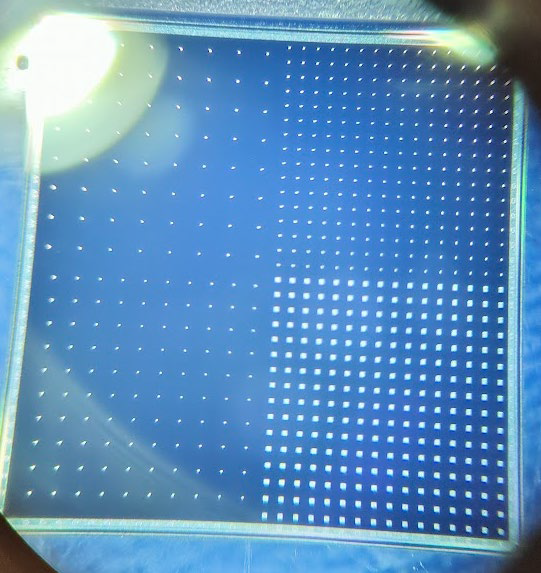}
 \includegraphics[width=0.4\columnwidth]{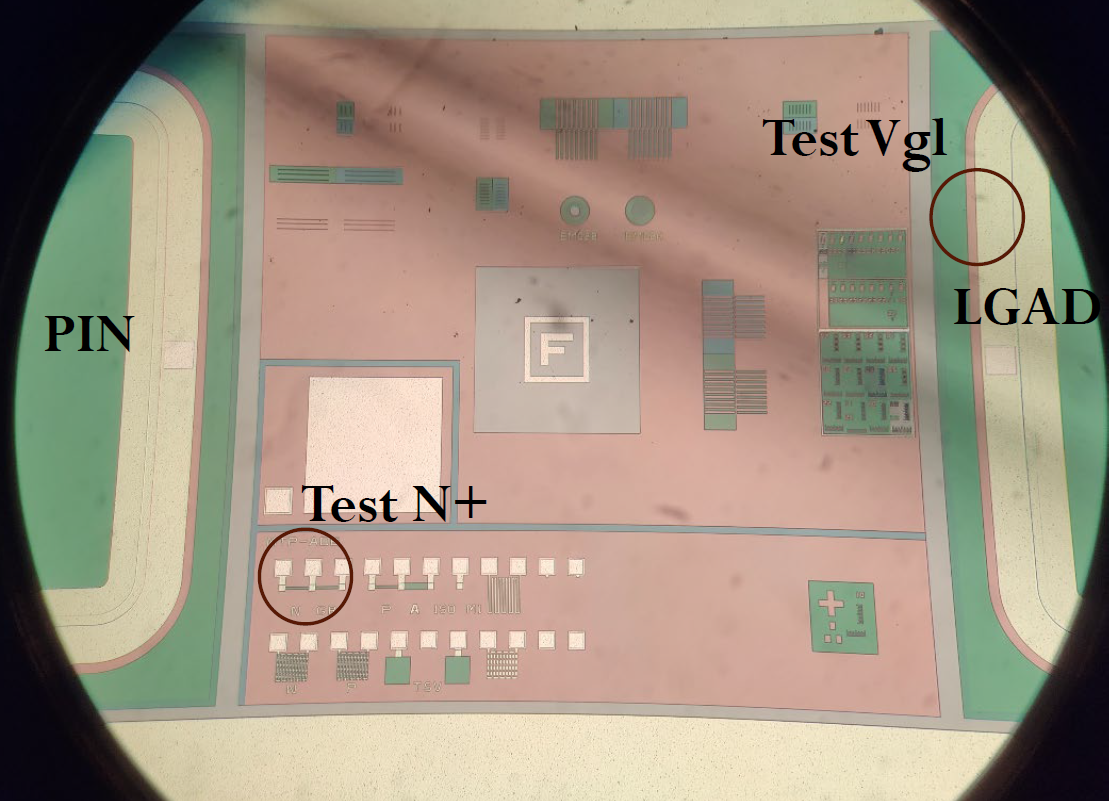}
  \caption{Picture of a full-size pixel sensor (Left) and test structures (Right).}
 \label{fig:ACLGAD}
\end{figure}

AC$-$LGADs were produced by Hamamatsu Photonics K.K. (HPK)~\cite{Hamamatsu} for ePIC and paid for by ePIC project funds. The TOF in ePIC is divided into barrel TOF, using 2x3~cm strip detectors (Fig.~\ref{fig:ACLGAD}, Left), and forward TOF, using 1.6x1.6~cm pixel detectors (Fig.~\ref{fig:ACLGAD}, Center). 
Each produced wafer also contained several test structures shown in Fig.~\ref{fig:ACLGAD}, Right; the structure contains a single pad PIN and LGAD to test gain layer degradation and a test point to measure the resistivity of the N+ resistive layer.
The characterization of the pre-irradiation devices was reported in a previous paper~\cite{davis2025characterizationfullsizeproductionepic}.
Several pixel detectors and test structures for pixels and strips were irradiated according to Tab.~\ref{tab:wafers}. Full-size strip detectors were not irradiated due to limited availability and size not fitting the standard irradiation vessel.
In addition, a few DC-LGADs from past production were irradiated, also reported in Tab.~\ref{tab:wafers}.
\begin{table}[ht]
\centering
\begin{tabular}{|l|lll|l|} 
\hline
Geometry & N+ ($\Omega/\square$) & Thick. ($\mu$m) & Size (cm) & Fluence ($n_{eq}/cm^2$)\\
\hline
Strips & 1600 & 50-30 & 3x2 & Not irradiated\\
Test structure & 1600 & 50-30 & - & 2e17, 5e17, 1e18\\
\hline
Pixels & 600 & 30-20 & 1.6x1.6&2e17, 5e17, 1e18\\
Test structure & 600 & 30-20 & - &2e17, 5e17, 1e18\\
\hline
\end{tabular}
\caption {Geometry and thickness of the tested HPK AC-LGAD wafers. Strip sensors were not irradiated due to limited availability.}
\label{tab:wafers}
\end{table}
The sensors were irradiated without bias in the JSI research reactor of TRIGA type in Ljubljana, which has been used successfully in the past decades to support the development of radiation-hard sensors~\cite{JSI_facility}.
The neutron energy spectrum and flux are well known. Fluences at JSI are between \fluence{2}{17} to \fluence{1}{18}. 
The fluence is quoted in 1~MeV equivalent neutrons per cm$^2$ ($n_{eq}/cm^2$) by using Non-Ionizing Energy Loss (NIEL) scaling.
The uncertainty of the fluence at the TRIGA reactor is around $5\%$.
After irradiation, the devices were annealed in steps of 80 minutes at 60~\degree C and measured after each step. The devices were kept in cold storage at -20~\degree C to reduce further annealing while not in testing.

\section{Laboratory setups}
Due to the high irradiation fluence of the sensors, measurements of current versus voltage (IV) were conducted in a climate chamber.
Sensors and test structures were mounted onto boards without active electronics and connected to a HV supply (Keithley 2657A).
Current versus voltage was measured to find the breakdown voltage, at which the leakage current increases exponentially.
The N+ layer resistivity of the test structures was measured in a probe station without temperature control capabilities (at room temperature), where a voltage and ground were applied through two tungsten needles on the test structure. 
IV measurements for pre-irradiation samples with low current were performed in a probe station with a chuck connected to an HV power supply and needles for ground connection.

The capacitance versus voltage (CV) was also measured both in the climate chamber and in the probe station with an additional connection to an LCR meter (Agilent E4980A) and a capacitor decoupling box. The CV box had an internal series resistor of 100~k$\Omega$ and a maximum voltage of $\pm$200~V. Due to the high leakage current of the tested samples, only a fraction of the bias voltage is applied to the sensors, and the correct voltage had to be found by calculating $V_{corr}$ = $V_{meas}$ - 100k$\Omega$$\times$I. To reduce this effect, several CVs were also conducted in the climate chamber at colder temperatures. The LCR meter frequencies were set to 1~kHz and 10~kHz in the room-temperature probe station, while the LCR meter frequencies were set to 100, 200, and 500~Hz when in the climate chamber. Capacitance versus voltage measurements were carried out to attempt to find the gain layer's depletion voltage $V_{GL}$, although it was not possible to get a measurement for the fluences tested.



The charge collection measurements using TCT follow the method described in~\cite{BISHOP2024169478}. In short, the sensors are mounted on fast analog amplifier boards with 16 channels and a bandwidth of 1~GHz designed at Fermilab (FNAL)~\cite{Heller:2022aug} and read out by a fast oscilloscope (2 GHz, 20 Gs). 
The laser station has a cold chuck with a Peltier cell and recirculating cooling that can maintain a temperature of $-$10~C around the sensor, monitored by a PT100 sensor soldered on the board.
An infrared (IR) 1064 nm pulsed laser with a pulse temporal width of 30 ps and a spot of 10-20~$\mu$m width mimics the response of a MIP in the silicon~\cite{Particular}. The IR laser cannot penetrate metal; therefore, the sensor behavior can be characterized only between metal electrodes.
The response of the sensor as a function of laser spot position is evaluated using X-Y moving stages. Waveforms are averaged for each position to decrease noise and the effect of laser power fluctuations. 
The pulse maximum (Pmax)~\cite{Ott:2022itj}, rise time (10-90\%), and time of arrival (time of the pulse maximum, Tmax) are calculated for each position and plotted as a function of position. The Pmax distributions shown are normalized to the maximum value of each respective distribution for ease of comparison. The laser power had to be increased to see a reasonable signal from the irradiated devices.

\section{Electrical results}

\subsection{Irradiation effects on sensors}
\begin{figure}[!hbt]
 \centering
 \includegraphics[width=0.48\columnwidth]{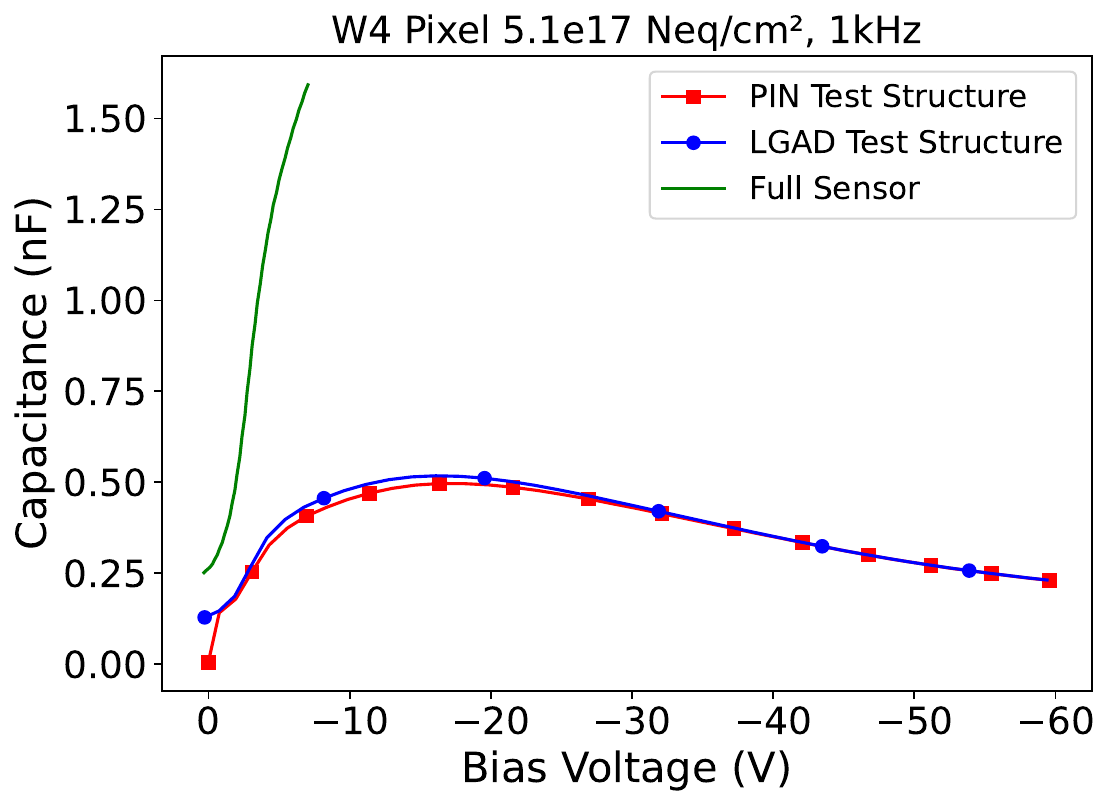}
  \includegraphics[width=0.48\columnwidth]{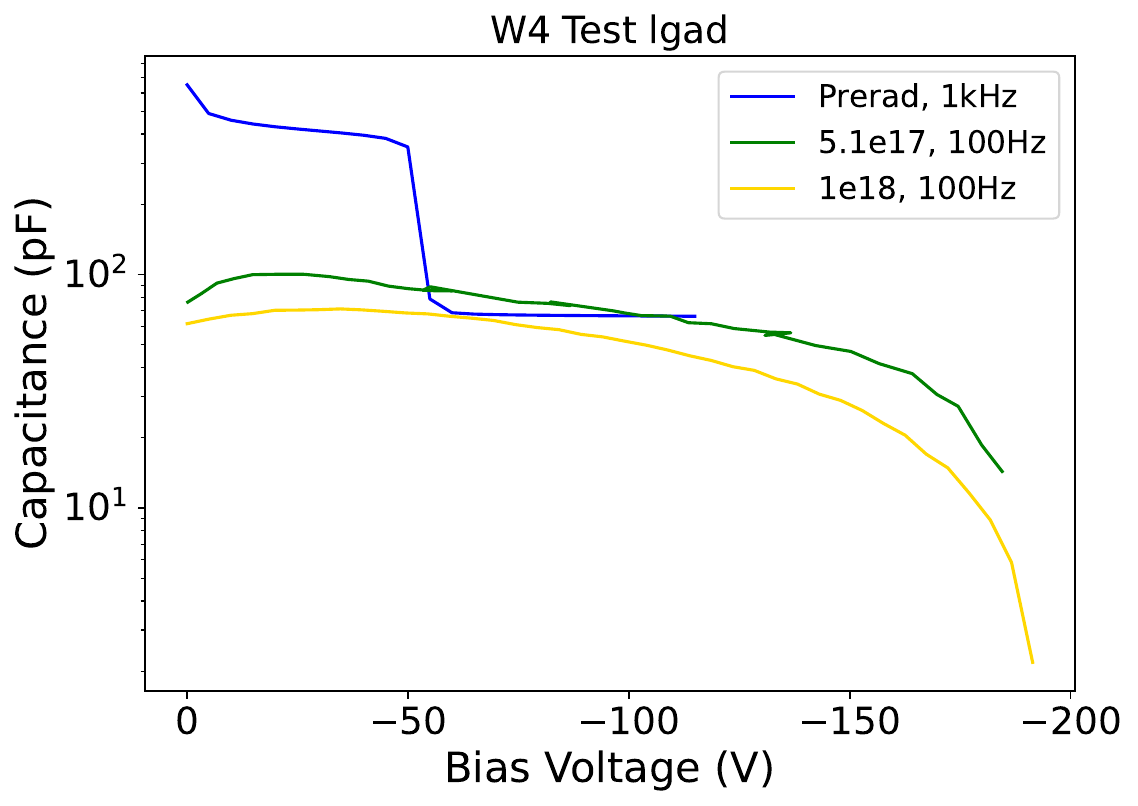}
  \caption{Left: CV after 2.3e17~$n_{eq}/cm^2$ of irradiation of PIN (left) and LGAD (right) pixel test structures plus a full-size pixel sensor (5,6). The leakage current for the full-size pixel is too high to allow a CV measurement over 10V. All measurements are at room temperature with 10kHz frequency.
  Right: Capacitance over voltage before and after irradiation of a pixel LGAD test structure. (Final Cap Pre-rad: W8 - 46.7pF, W4 - 66.3pF, W11 - 25.0pF)
  {\color{red} TODO: add explanation on thickness.}
  }
 \label{fig:cvPINLGAD}
\end{figure}

The variation of the gain layer depletion with irradiation can usually be observed until a fluence of 1e16~$n_{eq}/cm^2$, as shown in previous studies for traditional LGADs~\cite{Padilla_2020}. 
However, for the extreme fluences presented in this paper, it's not viable to get a $V_{GL}$ measurement. As shown in Fig.~\ref{fig:cvPINLGAD}, the PIN and LGAD structures exhibit the same behavior already at a fluence of 2.3e17~$n_{eq}/cm^2$, indicating that no trace of the gain layer is left in the LGAD devices.
In previous studies~\cite{Stage:2025jre}, the $V_{GL}$ vs fluence distribution from each wafer was fit with $N_D=N_0e^{-c\phi}$ to measure the \textit{c}-factor for each wafer. The \textit{c} values generally vary between 4.5 and 5.5. 

The capacitance vs. voltage behavior is different from usual, with capacitance increasing at first and then decreasing. This measurement was done at room temperature up to 200~V; however, the system has an in-line 100~$K\Omega$ resistor which drops the maximum voltage to around 60~V due to the high current. 

\begin{figure}[!hbt]
 \centering
 \includegraphics[width=0.7\columnwidth]{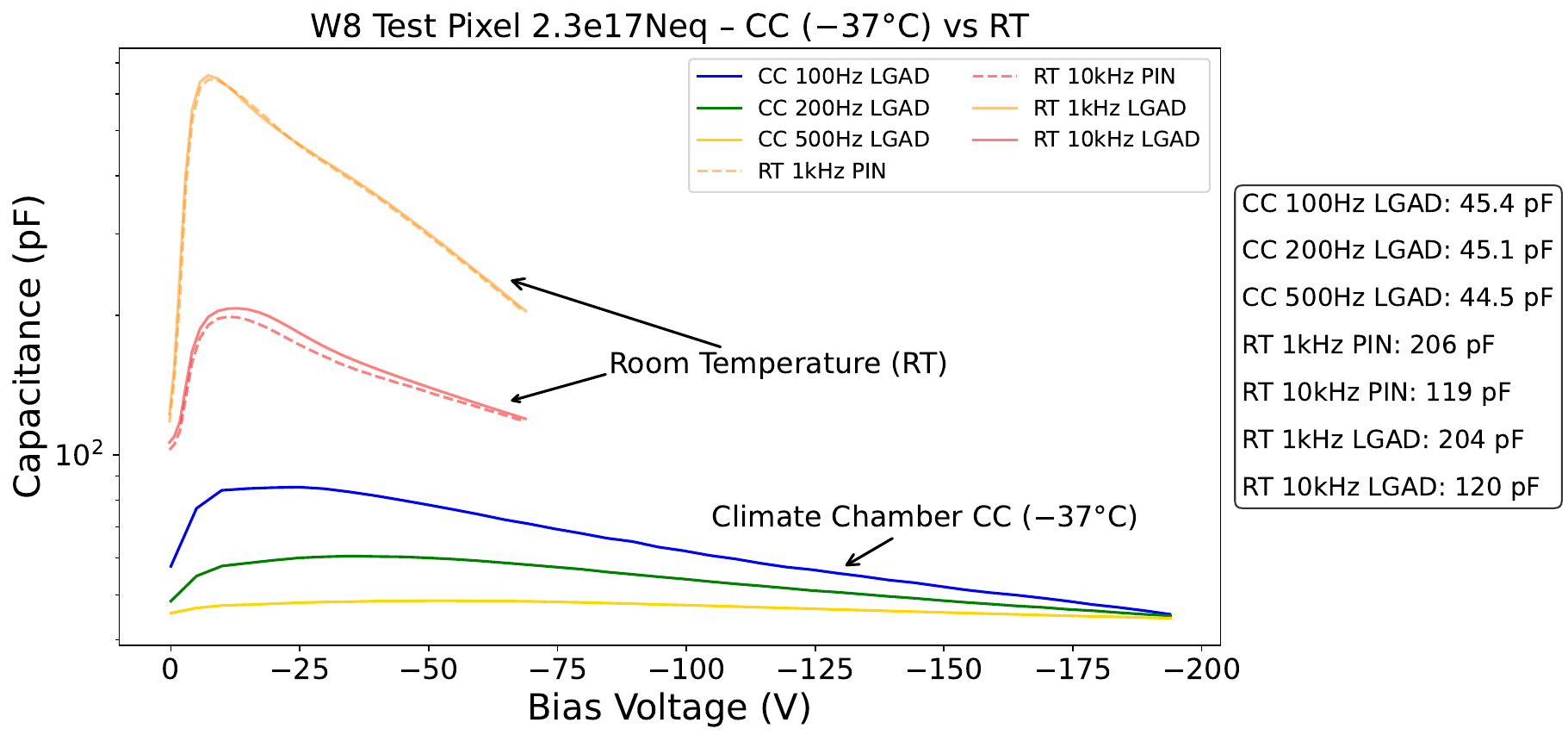}
 \caption{CV of an irradiated sensor at room temperature and at -37C in the climate chamber. 
 {\color{red} TODO: improve plot, simplify message, better legend, say it reaches full depletion}
 }
 \label{fig:cvcold}
\end{figure}

As shown in Fig.~\ref{fig:cvcold}, full depletion of the device cannot be achieved at room temperature with the current setup. However, it can be reached at a colder temperature (-37C) in a climate chamber, as the leakage current lowers with temperature. 
A dependence on the frequency employed in the CV was observed for the irradiated devices. However, the three frequencies roughly converge to the same final capacitance in the cold measurement. 
Before irradiation, the capacitance did not change with frequency.

\FloatBarrier

\subsection{Irradiation effects on N+ Charge Sharing Layer}
\begin{figure}[!hbt]
 \centering
 \includegraphics[width=0.45\columnwidth]{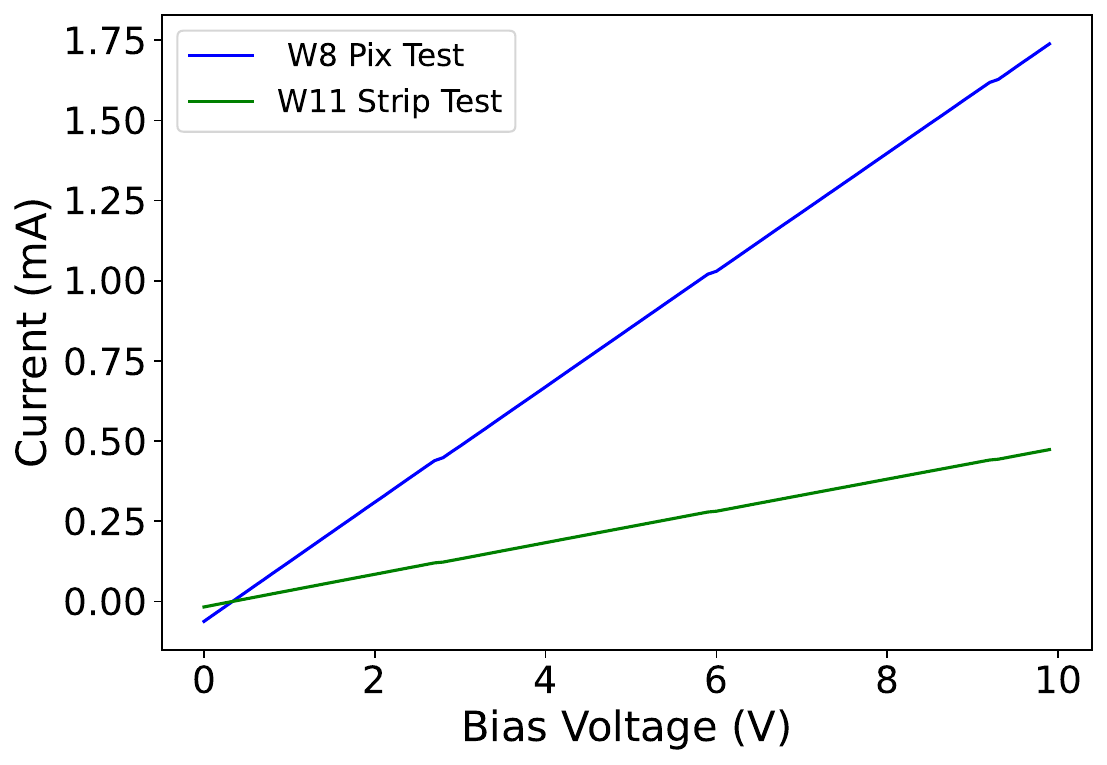}
 \includegraphics[width=0.45\columnwidth]{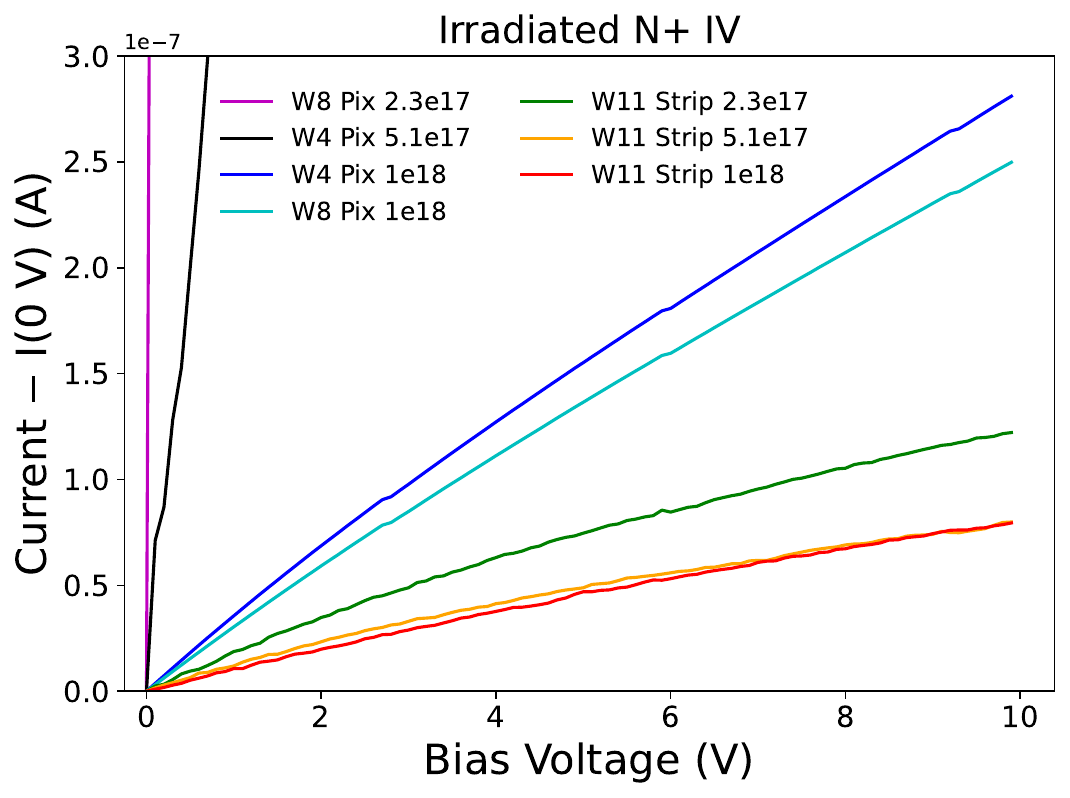}
  \caption{Left: N$^+$ sheet conductance of pre-irradiated test structures for pixel and strips corresponding to 560~$\Omega$ and ?~$\Omega$ respectively. Right: N$^+$ sheet conductance of irradiated test structures. The absolute current is shifted to compare the distributions. The minimum current is subtracted and 15\% of the range is added.
  \color{red} TODO: how is it shifted?
  }
 \label{fig:NplusFitfluence}
\end{figure}

The N+ resistivity was tested using the test structures in Fig.~\ref{fig:ACLGAD}, right. As reported in previous papers~\cite{ELICABUK2025170705}, the resistivity is expected to increase as the donors get deactivated. The resistivity of n-type silicon can be approximated using the formula:
\[
\rho \approx \frac{1}{q \cdot N_D \cdot \mu_n}
\]
\noindent Where, $\rho$ is the measured resistivity ($\Omega\cdot\text{cm}$), $q$ is the elementary charge ($1.602 \times 10^{-19}\text{ C}$), $N_D$ is the donor doping sheet concentration ($\text{cm}^{-3}$), $\mu_n$ is the electron mobility ($\text{cm}^2/\text{V}\cdot\text{s}$). The active donor concentration is proportional to the conductivity, which is $1/\rho$.
This is observed for both pixels and strips, as shown in Fig.~\ref{fig:NplusFitfluence}. The N+ resistivity for the pixel-type detector goes from a 600~$\Omega/\square$ sheet up to 13000~$\Omega/\square$ of sheet resistance.
The $N^+$ resistivity affects the device's resistive behavior and the charge sharing between electrodes, as reported in~\cite{BISHOP2024169478}. Therefore, a large increase will be reflected in the sensor's charge collection behavior, as shown in the following section.
Don't know thickness so can't get doping

{\color{red} TODO: get the doping from the resistivity and fit the donor removal coefficient.}
{\color{red} TODO: try to do the same for P test structure.}

\begin{figure}[!hbt]
 \centering
 \includegraphics[width=0.45\columnwidth]{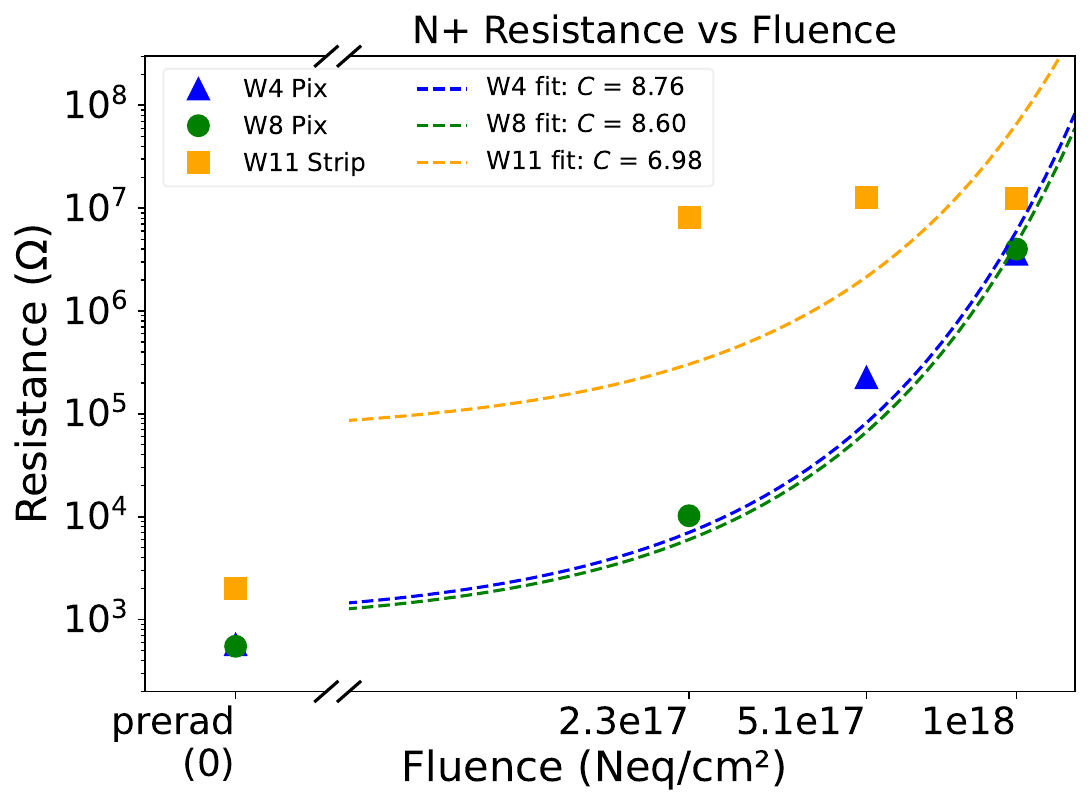}
 \includegraphics[width=0.45\columnwidth]{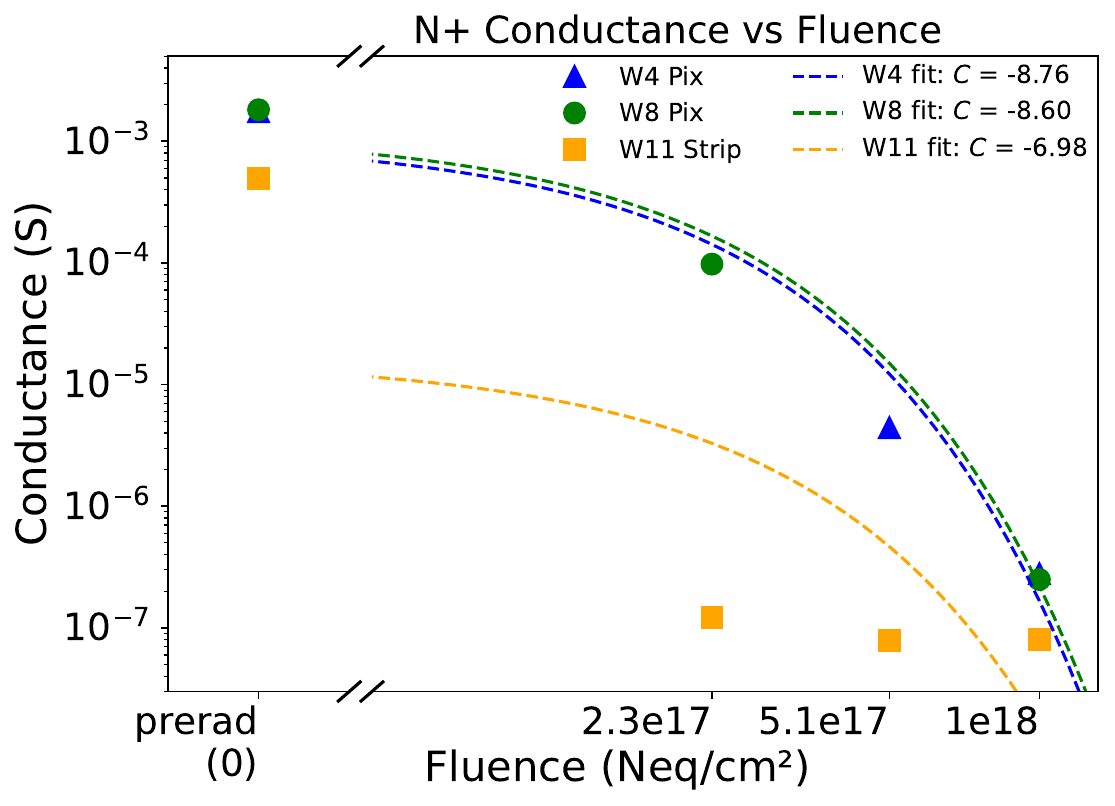}
  \caption{N$^+$ sheet conductance from test structure for pixel test structures, the pre-irradiation value is shown at 1e12~Neq to simplify visualization. \color{red} TODO: move the prerad to higher fluence? Show fit?}
  \label{fig:NplusFluence}
\end{figure}

\subsection{Annealing effects}
\begin{figure}[!hbt]
 \centering
 \includegraphics[width=0.55\columnwidth]{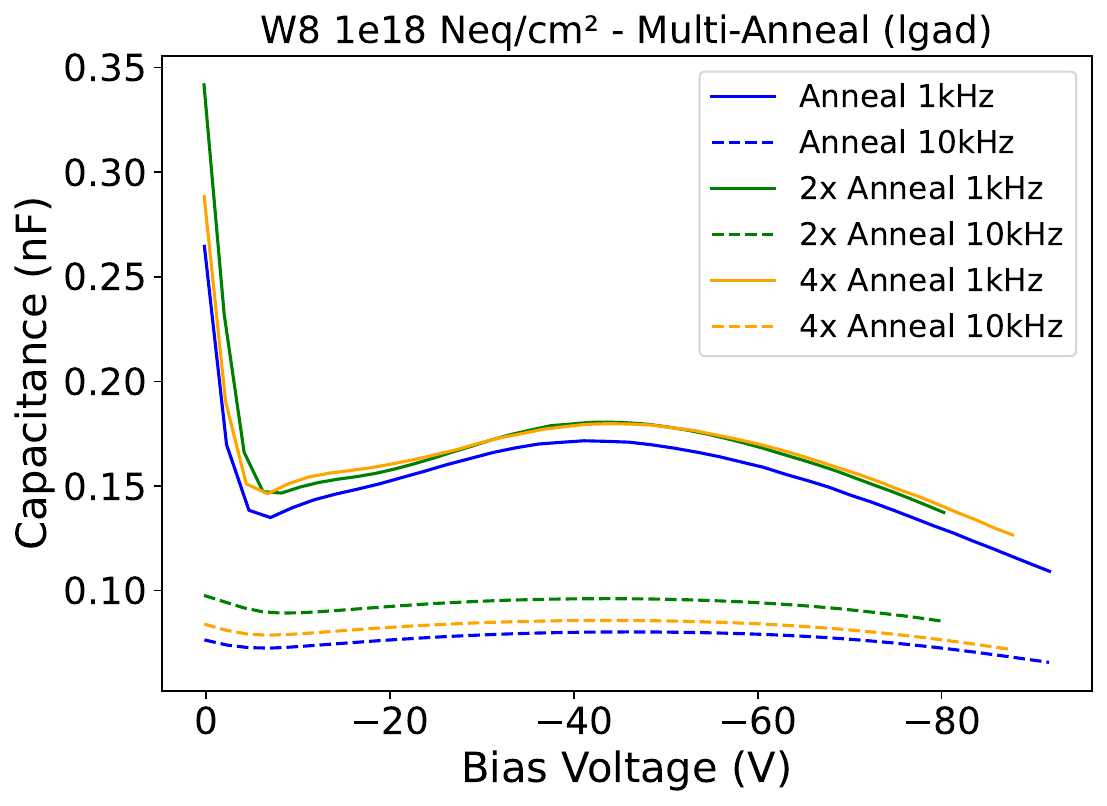}
  \caption{ CV measurement of a test strucutre irradiated at 1e18Neq after different annealing cycles. The measurements was done at 1KHz and 10KHz. 
  }
  \label{fig:annealing}
\end{figure}

The samples were annealed before testing with a thermal treatment of 60C for 80 minutes in a temperature-controlled climate chamber. The samples were tested, and afterwards the same annealing procedure was repeated to reach a total of 160 minutes annealing. After further testing, the samples were annealed at the same temperature but for 160 minutes, to reach a total of four annealing cycles with 340 minutes of annealing. As seen in Fig.~\ref{fig:annealing}, multiple annealing do not have minor effects to both the CV measurement or N+ resistivity.

\begin{figure}[!hbt]
 \centering
  \includegraphics[width=0.45\columnwidth]{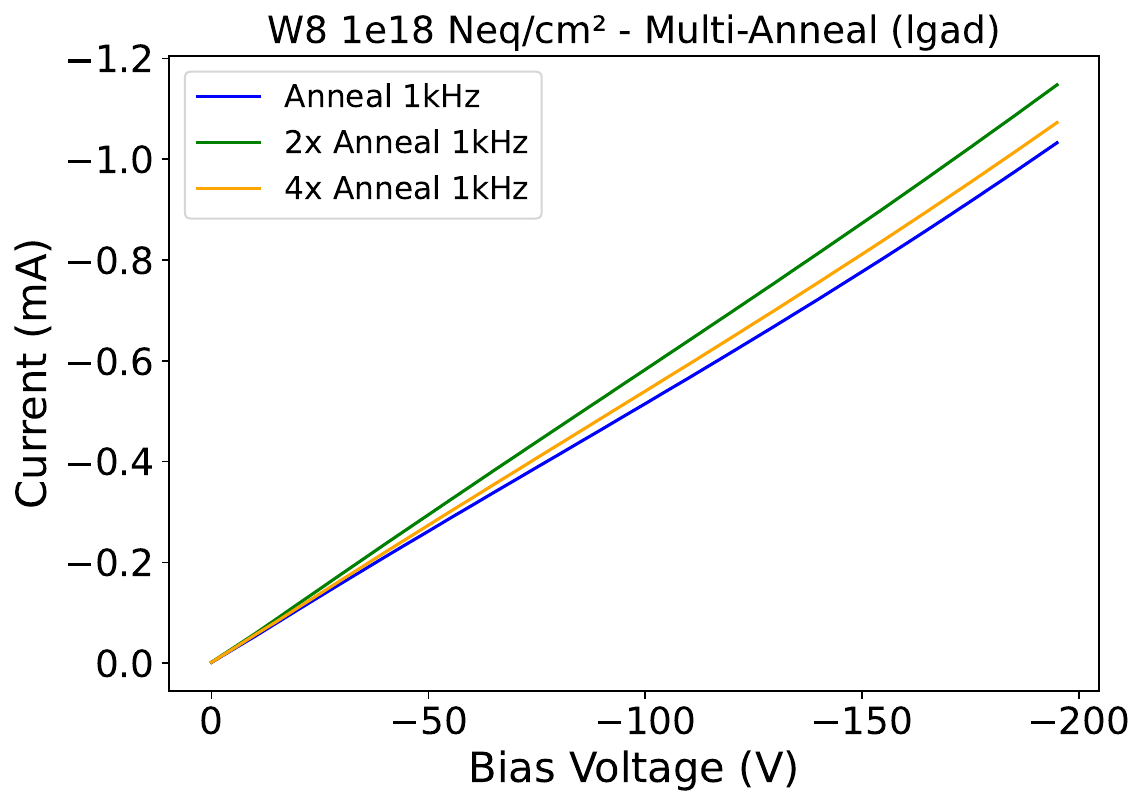}
 \includegraphics[width=0.45\columnwidth]{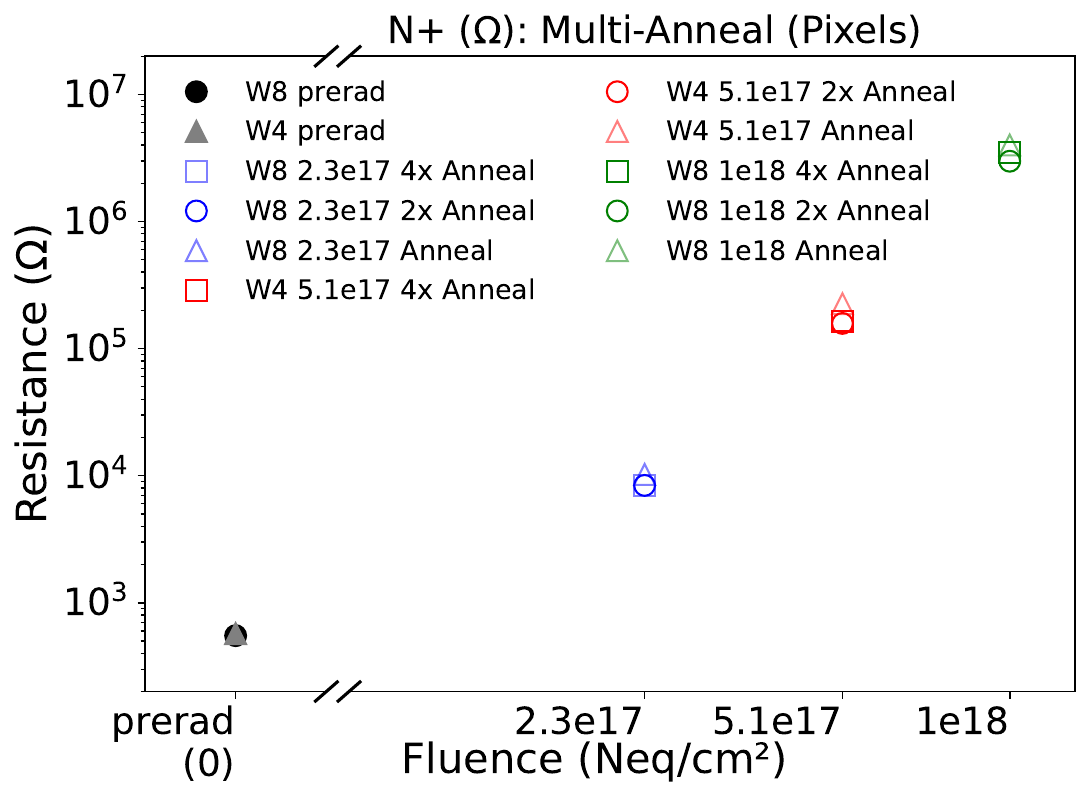}
  \caption{Left: N$^+$ sheet resistivity from test structure for pixel (560~$\Omega$ pre-irradiation value) test structures after multiple annealing cycles. Right: N$^+$ sheet conductance from test structure for pixel (560~$\Omega$ pre-irradiation value) teststructures. \color{red} TODO: why no 4x anneal?
  }
  \label{fig:NplusFluenceAnneal}
\end{figure}

\section{Laser TCT results}
A selection of irradiated pixel detectors were tested with the laser TCT in the station described in the previous section. The response as a function of position for the large pixel devices was probed for all irradiation levels and compared with the behavior of a non-irradiated sensor. 
The variables calculated from the waveforms are the following: Pmax, maximum of the pulse, Rise Time, 10-90\% rise time of the pulse, and CFD50, time of arrival from the trigger taken as the 50\% point of the pulse.

\begin{figure}[!hbt]
 \centering
  \includegraphics[width=0.35\columnwidth]{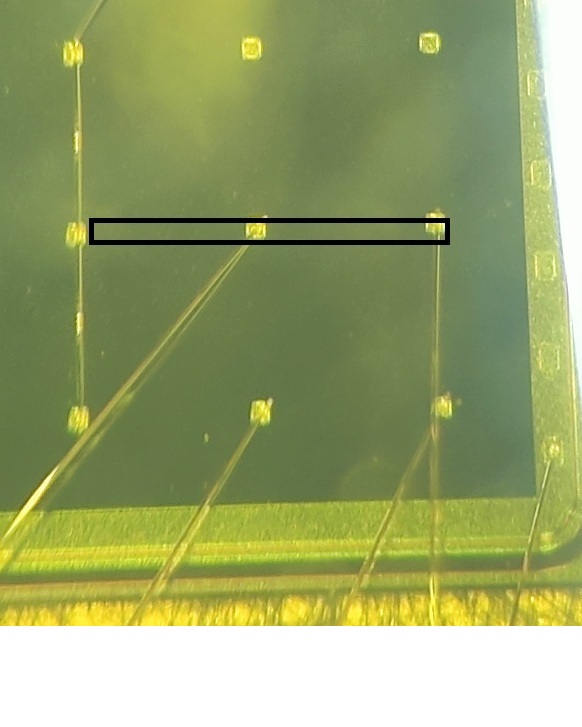}
 \includegraphics[width=0.64\columnwidth]{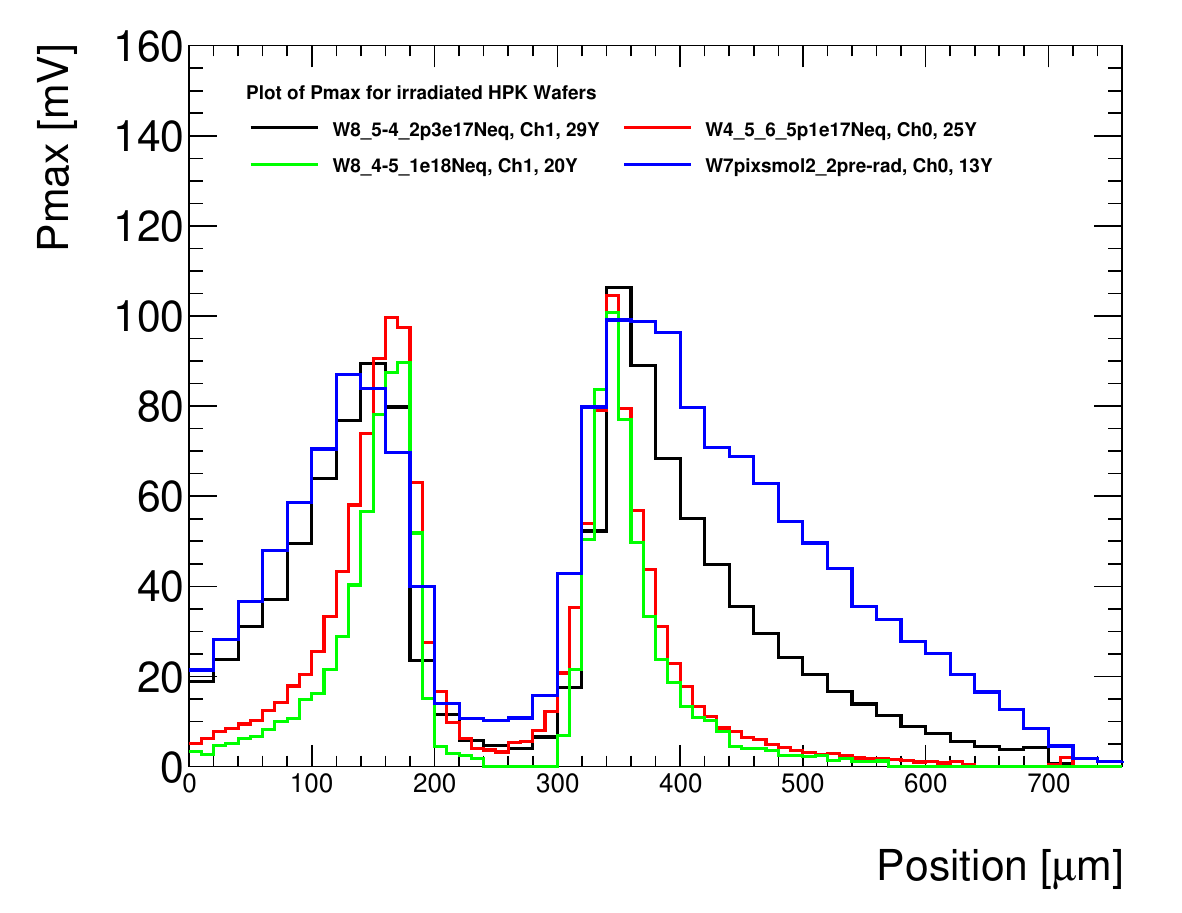}
  \caption{Left: area scanned with the laser TCT. Right: Normalized Pmax versus position for before- and after-irradiation devices.}
 \label{fig:Pmaxfluence}
\end{figure}

\begin{figure}[!hbt]
 \centering
  \includegraphics[width=0.4\columnwidth]{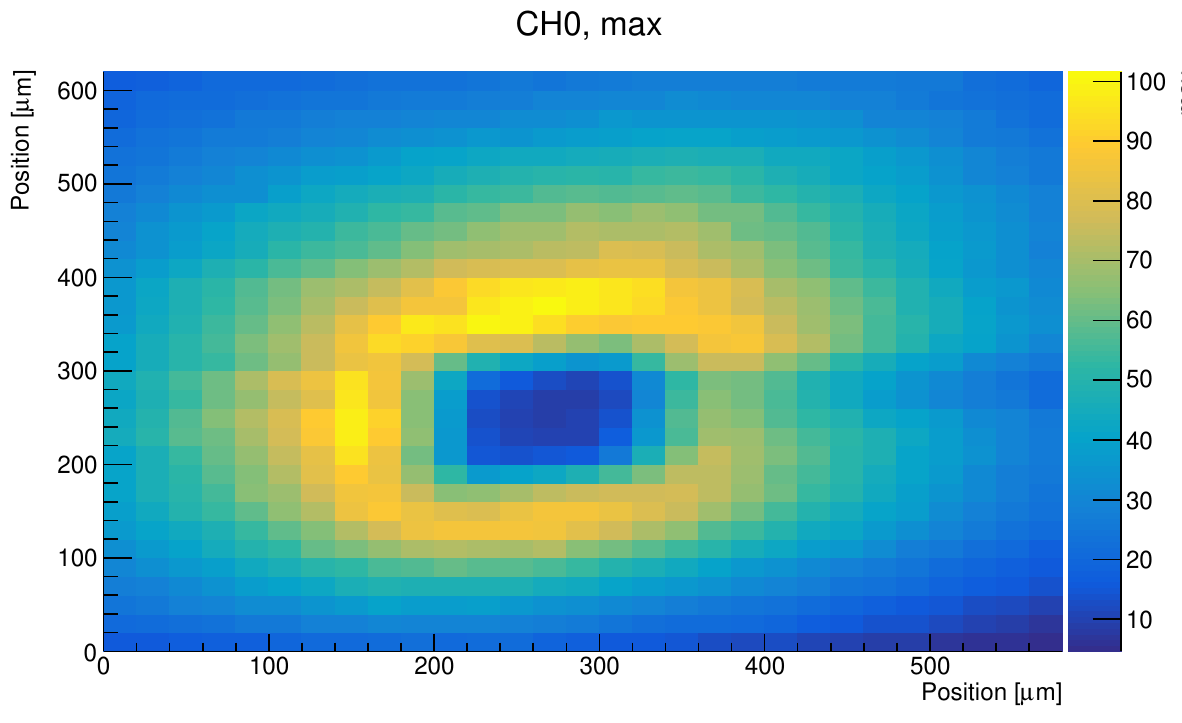}
 \includegraphics[width=0.4\columnwidth]{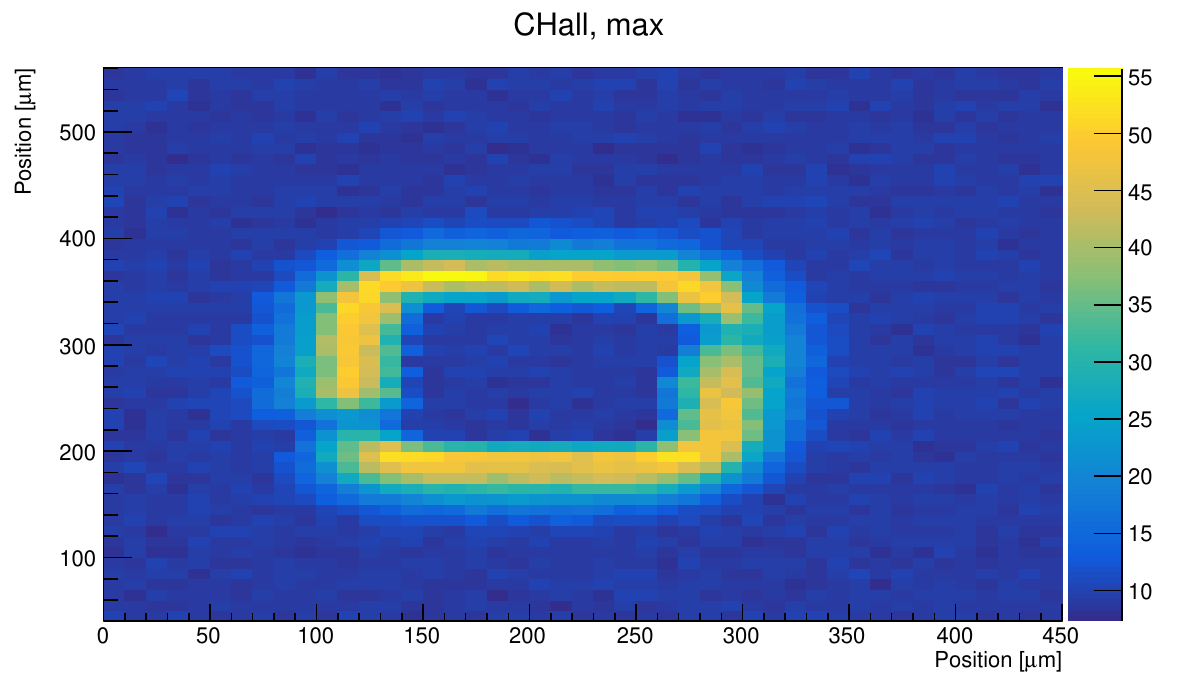}
  \caption{2D area scanned with the laser TCT for non-irradiated (right) and irradiated at 1e18Neq (left).}
 \label{fig:Pmaxfluence2D}
\end{figure}

Fig.~\ref{fig:Pmaxfluence} shows the response as a function of position for a single pixel in the large sensor. Fig.~\ref{fig:Pmaxfluence2D} is a 2D scan around one pixel for non-irradiated and irradiated sensors. The response is normalized, as the post-irradiation has a significantly lower signal. The laser is tuned to a 1 Minimum Ionizing Particle for the before-irradiation device but had to be increased in intensity to see a good signal on the irradiated devices {\color{red} what's the laser power? prerad - 100, 2.3e17 - 325, 5.1e17 - 400, 1e18 - 300}, which were biased around 200V.
As expected, the non-irradiated device with the lowest $N^+$ resistivity has the larger fraction of signal vs distance. The irradiated device shows a gradual reduction in the charge sharing profile, which is explained by the increased resistivity. TCAD simulations of charge sharing profiles as a function of resistivity can be found in~\cite{Nizam:2023zgg}; studies in~\cite{BISHOP2024169478} confirmed the same behavior.

\begin{figure}[!hbt]
 \centering
 \includegraphics[width=0.45\columnwidth]{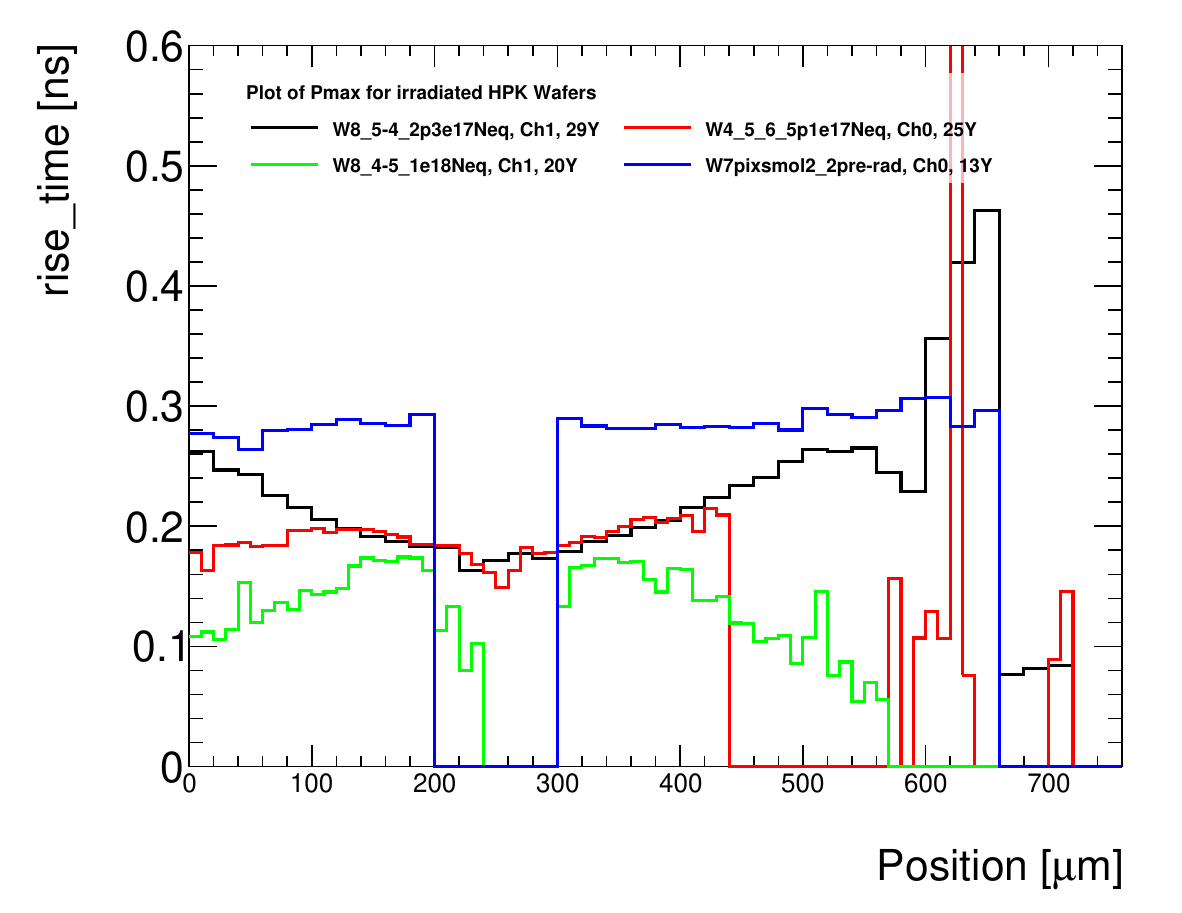}
 \includegraphics[width=0.45\columnwidth]{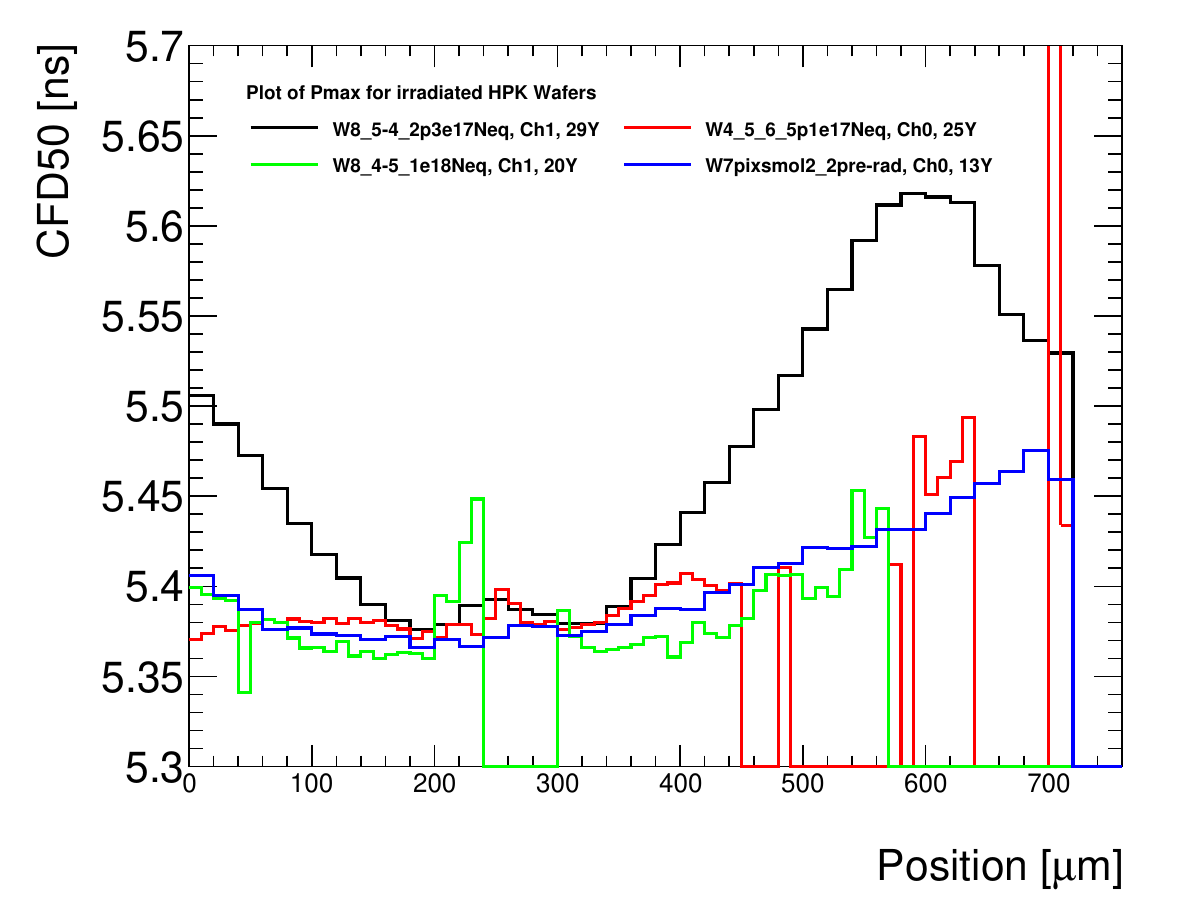}
  \caption{Left: Rise time version position for before and after irradiation devices. Right: Time of arrival (CFD50)  {\color{red} TODO: recheck 2.3e17Neq. Make the plot more understandable.}}
 \label{fig:timefluence}
\end{figure}

Fig.~\ref{fig:timefluence} shows the rise time and CFD50\% as a function of position in the same region. The rise time (Fig.~\ref{fig:timefluence}, Left) of 300~ps is expected for the non-irradiated sensors, as it's a 30um thick device and the rise time depends on the drift time of the electrons to the surface for the multiplication mechanism to happen. The rise time is reduced for irradiated sensors, as no gain is left and the pulse is instantaneous, as for standard silicon PIN detector.
The CFD50 (Fig.~\ref{fig:timefluence}, Right) highlights that the travel time of the signal in the $N^+$ is roughly unchanged with resistivity.

\begin{figure}[!hbt]
 \centering
 \includegraphics[width=0.45\columnwidth]{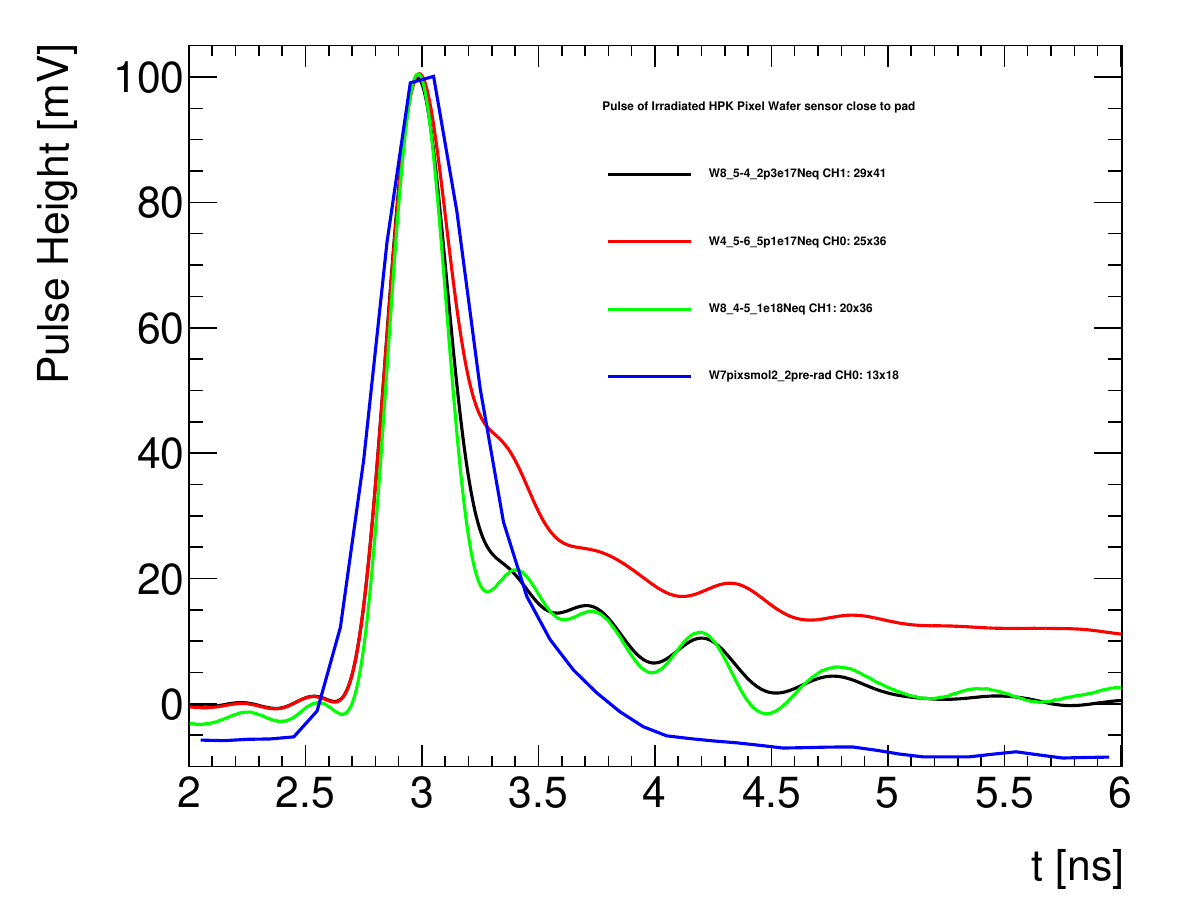}
 \includegraphics[width=0.45\columnwidth]{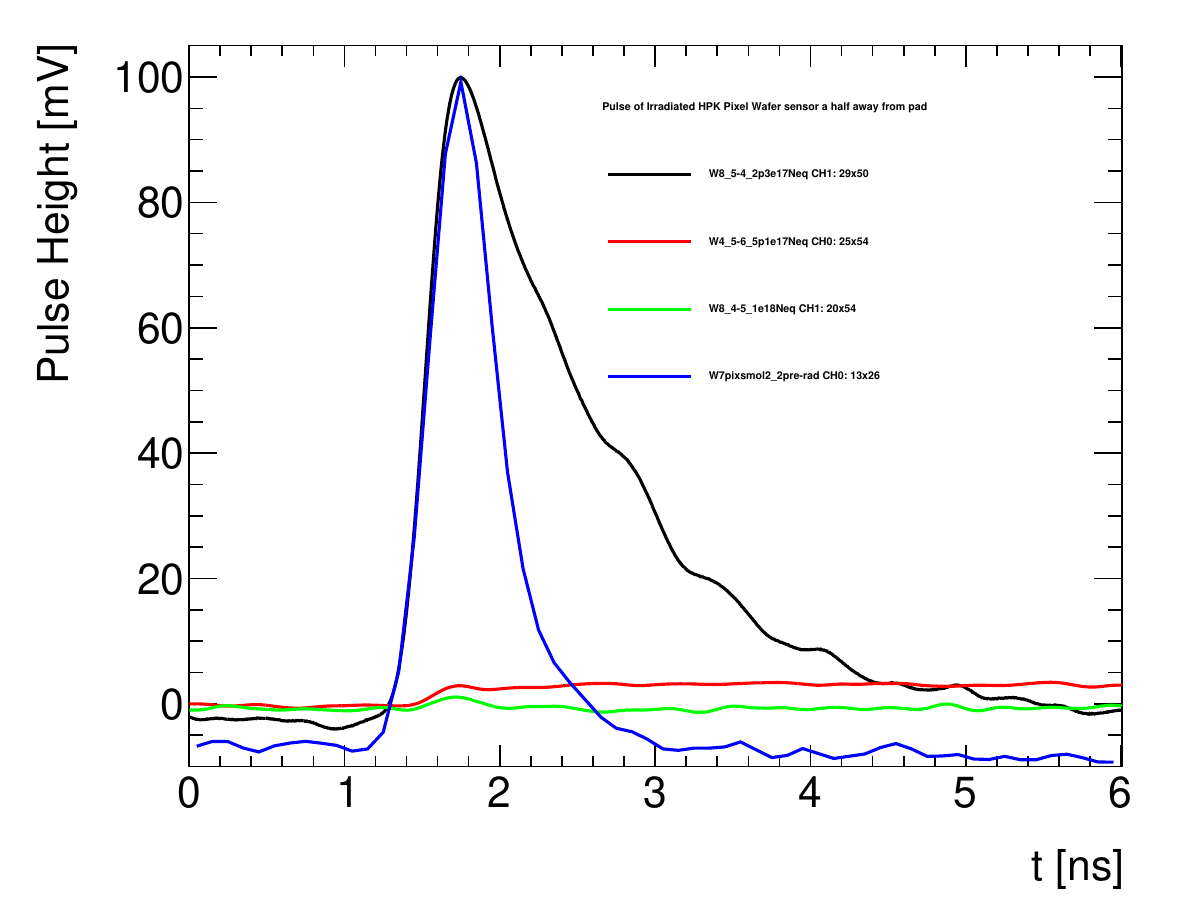}
  \caption{Average waveform for the laser next to the pixel (Left) and halfway with the next neighbor pixel (Right) {\color{red} TODO: keep normalization of 2.3e17Neq to plot on the left.}}
 \label{fig:waveforms}
\end{figure}

The averaged waveforms (of 100 samples) are shown in Fig.~\ref{fig:waveforms} next to the pixel (Left, around 350um in Fig.~\ref{fig:Pmaxfluence}) and between two pixels (Right, around 550um in Fig.~\ref{fig:Pmaxfluence}). The pulses are normalized as in Fig.~\ref{fig:waveforms}. The device before irradiation show an increased rise time as expected from Fig.~\ref{fig:timefluence}. Almost no signal is present after heavy irradiation in between the pixels.

\section{Conclusions}
AC-LGADs and test structures from the same wafers produced by the ePIC collaboration were irradiated at extreme fluences (2e17 to 1e18 Neq) at the JSI Triga reactor. The sensors were characterized post-irradiation with a probe station and a laser TCT station. The behavior was compared with the same devices before irradiation. After the lowest irradiation point (2e17Neq), no trace of a gain layer is left, and the LGADs are indistinguishable from PIN detectors. 
Due to the elevated leakage current, the sensors can only be operated successfully in a cold environment. 
Overall, no annealing effect was observed in the sensors after the first round of annealing at 80 minutes at 60~C.

The N+ resistive layer of the AC-LGADs was tested using test structures present on the wafer. The resistivity increases from O(1~k$\Omega$) to O(1~M$\Omega$) due to the large radiation damage. 
The increased resistivity affects the sensor response; the metal electrodes present increased isolation at 1e18~Neq, showing a response profile of the pixel falling to <10\% of the signal under the electrode at 100~$\mu m$ of distance from the metal, instead of a 500~$\mu m$ distance for non-irradiated device.
The effect is gradual, showing a intermediate behavior at 2.3e17~Neq and plateaus at 5e17~Neq which presents the same behavior of 1e18~Neq.
An increased laser intensity was necessary to observe a good signal from the irradiated devices. The loss of gain causes the signal to have an instantaneous rise time, which is compatible with a PIN detector behavior. 
The signal propagation speed in the N+ is not affected by the irradiation as observed in previous studies~\cite{Stage:2025jre}.

\section{Acknowledgments}
We thank the technicians and students at SCIPP for the support and the time spent in the lab.
The sensors were produced with funds from the ePIC erd112 effort for AC-LGADs development for the ePIC detector TOF layer.
The irradiation at the TRIGA reactor was covered by the European Union’s Horizon Europe Research and Innovation program under Grant Agreement No 101057511 (EURO-LABS) in the context of the DRD3 collaboration.
This work was supported by the United States Department of Energy grants DE-SC0010107.

\bibliography{bib/TechnicalProposal,bib/hpk_fbk_paper,bib/HGTD_TDR,bib/SHIN,bib/nizam,bib/others}

\end{document}